\documentclass[aps,pre,reprint,twocolumn]{revtex4-1}

\usepackage{graphicx}% Include figure files
\usepackage{mathtools}
\usepackage{tikz}
\usepackage{comment}
\usepackage[caption=false]{subfig}
\usepackage[colorlinks,allcolors=blue]{hyperref}% add hypertext capabilities

\begin{document}

% Use the \preprint command to place your local institutional report
% number in the upper righthand corner of the title page in preprint mode.
% Multiple \preprint commands are allowed.
% Use the 'preprintnumbers' class option to override journal defaults
% to display numbers if necessary
%\preprint{}

%Title of paper
\title{On the emergence of a power law in the distribution of COVID-19 cases}

% repeat the \author .. \affiliation  etc. as needed
% \email, \thanks, \homepage, \altaffiliation all apply to the current
% author. Explanatory text should go in the []'s, actual e-mail
% address or url should go in the {}'s for \email and \homepage.
% Please use the appropriate macro foreach each type of information

% \affiliation command applies to all authors since the last
% \affiliation command. The \affiliation command should follow the
% other information
% \affiliation can be followed by \email, \homepage, \thanks as well.
\author{Brendan K.\ Beare}
\email{brendan.beare@sydney.edu.au}
%\homepage[]{Your web page}
%\thanks{}
%\altaffiliation{}
\affiliation{School of Economics, University of Sydney, Sydney, NSW 2006, Australia}

\author{Alexis Akira Toda}
\email{atoda@ucsd.edu}
%\homepage[]{Your web page}
%\thanks{}
%\altaffiliation{}
\affiliation{Department of Economics, University of California San Diego, La Jolla, CA 92093, USA}

%Collaboration name if desired (requires use of superscriptaddress
%option in \documentclass). \noaffiliation is required (may also be
%used with the \author command).
%\collaboration can be followed by \email, \homepage, \thanks as well.
%\collaboration{}
%\noaffiliation

\date{\today}

\begin{abstract}
The first confirmed case of Coronavirus Disease 2019 (COVID-19) in the US was reported on January 21, 2020. By the end of March, 2020, there were more than 180000 confirmed cases in the US, distributed across more than 2000 counties. We find that the right tail of this distribution exhibits a power law, with Pareto exponent close to one. We investigate whether a simple model of the growth of COVID-19 cases involving Gibrat's law can explain the emergence of this power law. The model is calibrated to match (i) the growth rates of confirmed cases, and (ii) the varying lengths of time during which COVID-19 had been present within each county. Thus calibrated, the model generates a power law with Pareto exponent nearly exactly equal to the exponent estimated directly from the distribution of confirmed cases across counties at the end of March.
\end{abstract}

% insert suggested keywords - APS authors don't need to do this
%\keywords{}

%\maketitle must follow title, authors, abstract, and keywords
\maketitle

% body of paper here - Use proper section commands
% References should be done using the \cite, \ref, and \label commands
\section{\label{intro}Introduction}

In the natural and social sciences, a variety of size distributions exhibit a power law in the right tail, meaning that the fraction of observations whose size $S$ exceeds a threshold $s$ decays like a power function as that threshold increases: $\Pr(S>s)\sim s^{-\zeta}$ for large $s$. The parameter $\zeta>0$ is called the Pareto (or power law) exponent. Examples of distributions exhibiting power laws include income \cite{reed2003,Toda2011PRE,Toda2012JEBO,IbragimovIbragimov2018}, wealth \cite{KlassBihamLevyMalcaiSolomon2006,Vermeulen2018}, consumption \cite{TodaWalsh2015JPE,Toda2017MD}, city populations \citep{gabaix1999,reed2002,Soo2005,giesen-zimmermann-suedekum2010}, firm size \citep{Axtell2001,Fujiwara2004a}, family names \cite{MiyazimaLeeNagamineMiyajima2000,ZanetteManrubia2001,ReedHughes2002,ReedHughes2003}, stock returns \citep{GabaixGopikrishnanPlerouStanley2003,GabaixGopikrishnanPlerouStanley2006,GuIbragimov2018}, and numerous others \cite{mitzenmacher2004a,newman2005,gabaix2009,Eliazar2020}.

A common feature of models that purport to explain the prevalence of power laws is the presence of random multiplicative growth. Loosely, the size $S_t$ of a quantity of interest at time $t$ (e.g., wealth, population, firm size, etc.) is said to exhibit random multiplicative growth if its (random) growth factor $G_{t+1}= S_{t+1}/S_t$ is independent of the current size $S_t$. This is known among economists as Gibrat's law of proportional growth. On its own, Gibrat's law is not sufficient to generate a power law. For instance, a geometric Brownian motion (the continuous-time analogue of a random multiplicative growth process with lognormal growth) stopped at a fixed time has a lognormal distribution, which does not exhibit a power law. On the other hand, a geometric Brownian motion stopped at an exponentially distributed time has a double Pareto distribution, which exhibits a power law in both tails \cite{reed2001}. We may thus expect to observe a power law in the size distribution of a population whose members have been growing like geometric Brownian motions since birth, and whose distribution of ages is exponential. The combination of Gibrat's law with an exponential age distribution as a generative mechanism for power laws has been used extensively in recent economics literature \cite{luttmer2007,nirei-souma2007,benhabib-bisin-zhu2011,Toda2014JET,TodaWalsh2015JPE,AcemogluCao2015,Arkolakis2016,benhabib-bisin-zhu2016,GabaixLasryLionsMoll2016,AokiNirei2017,Toda2017MD,TodaWalsh2017JAE,CaoLuo2017,MukoyamaOsotimehin2019,Toda2019JME,StachurskiToda2019JET,MaStachurskiToda2020JET}. Related techniques have also been employed in the physics literature \cite{ManrubiaZanette1999,MonteroVillarroel2013,EvansMajumdar2014,MonteroVillarroel2016}.%; in the latter literature, the mechanism is typically framed as a logarithmic random walk affected by reset events, which can amount to the same thing.}%recent economic literature 

In this paper we study the distribution of confirmed cases of Coronavirus Disease 2019 (COVID-19) across US counties. As we will see, by the end of March 2020, a power law had emerged in the right tail of that distribution. We investigate whether the combination of Gibrat's law (for growth in the number of cases within a county) and a suitable age distribution (for the length of time since the outbreak in each county) can explain this power law. Using daily county-level data from the onset of COVID-19 in the US in January 2020 until the end of March 2020, we estimate the distributions of growth rates and ages, employing a gamma parametrization of the former and a truncated logistic parametrization of the latter. Our primary finding is that the Pareto exponent implied by the estimated growth rate and age distributions, which is 0.936, nearly exactly matches the Pareto exponent estimated directly from the distribution of cases across counties at the end of March, which is 0.930. This indicates that the combination of Gibrat's law with a suitable age distribution can explain the power law observed in the right tail of the distribution of COVID-19 cases across US counties.

A nice aspect of the COVID-19 data we analyze is that they span the entire history of confirmed cases in the US population, thus permitting us to observe the distribution of ages (days since outbreak) across counties. While there is limited empirical evidence that the age distributions of cities \cite{GiesenSuedekum2014} and firms \cite{Coad2010} may be approximately exponential, it is rarely the case that data used in economics and related fields allow a reliable estimate of the relevant age distribution. Indeed, it can be difficult to unambiguously define the age of a city, firm, or household, the latter of which is frequently interpreted as a dynastic unit spanning multiple generations. This confounds validation of the dynamic generative mechanism. Conveniently, our COVID-19 data reveal the entire shape of the age distribution, which allows us to provide what appears to be the first empirical analysis in which a Pareto exponent is obtained from direct estimates of both the growth rate and age distributions.

The remainder of our paper is organized as follows. Sec.~\ref{theory} contains theoretical background material. In Sec.~\ref{theory1} we explain how the combination of Gibrat's law and an exponential age distribution determines a Pareto exponent. In Sec.~\ref{theory2} we discuss the connection between Gibrat's law and a simple model of epidemics. Sec.~\ref{empirics} contains our empirical findings. In Sec.~\ref{empirics1} we describe our dataset. In Sec.~\ref{empirics2} we display the distribution of COVID-19 cases across US counties at the end of March, and report a Pareto exponent estimated directly from this distribution. In Sec.~\ref{empirics3} we assess the empirical plausibility of COVID-19 cases evolving according to Gibrat's law. In Secs.\ \ref{empirics4} and \ref{empirics5} we report our estimates for the distributions of growth rates and ages respectively. In Sec.~\ref{empirics6} we show how to compute the Pareto exponent implied by those growth rate and age distributions, and observe that it is close to the Pareto exponent reported in Sec.~\ref{empirics2}. Sec.~\ref{conclusion} contains brief remarks in nontechnical language summarizing the practical import of our findings. Our data and replication files are available online \footnote{\url{https://github.com/alexisakira/COVID-19_power_law}}.

\section{\label{theory}Theoretical background}

\subsection{\label{theory1}Power laws via Gibrat's law and exponentially distributed age}

Suppose that a unit (say, a county) starts with initial size (number of COVID-19 cases) $S_0=1$ at $t=0$ and grows randomly according to Gibrat's law of motion $S_t=G_tS_{t-1}$ for integer-valued $t\ge 1$, where $\{G_t\}_{t=1}^\infty$ is a sequence of independent and identically distributed (i.i.d.) positive random variables. Let $T$ be a random integer-valued time (days since COVID-19 outbreak in the county), independent of the sequence $\{G_t\}_{t=1}^\infty$, at which the unit size $S_T$ is observed. Suppose for now that $T$ has the geometric distribution (i.e., the discrete-time analogue of the exponential distribution), meaning that for $n\geq1$ we have $\Pr(T=n)=p(1-p)^{n-1}$ for some parameter $p\in(0,1)$ called a success probability.

What can be said about the right tail of the distribution of $S_T$? It turns out that, under a regularity condition on the distribution of the growth rate $X_t=\ln G_t$, the tail exhibits a power law. Specifically, letting $\mathrm{E}$ denote the expected value operator, we shall assume that the distribution of $X_t$ has Laplace transform $M(z)=\mathrm{E}(e^{zX_t})$ finite for real $z\in[0,\eta)$ and diverging to infinity as $z$ increases to $\eta$, where $\eta$ may be any positive real number or $+\infty$. Loosely, this means that $X_t$ can take positive values and has a distribution with a right tail that decays to zero exponentially or faster. When this regularity condition is satisfied, we may argue as follows to establish that the right tail of the distribution of $S_T$ exhibits a power law. Let $Y=\ln S_T=\sum_{t=1}^TX_t$. Observe that the distribution of $Y$ has Laplace transform $M_Y$ satisfying
\begin{equation}
M_Y(z)=\sum_{n=1}^\infty p(1-p)^{n-1}M(z)^n=\frac{pM(z)}{1-(1-p)M(z)}\label{eq:MY}
\end{equation}
for all positive real $z$ such that $(1-p)M(z)<1$. Noting that $M(z)$ is convex as a function of $z\in(0,\eta)$ and satisfies $M(0)<1/(1-p)<M(\eta)$, we deduce that there is a unique $\zeta\in(0,\eta)$ at which
\begin{equation}
(1-p)M(\zeta)=1,\label{eq:BeareToda}
\end{equation}
and that $M(z)$ has strictly positive derivative at $z=\zeta$. It thus follows from Eq.~\eqref{eq:MY} and an application of l'H\^{o}pital's rule that
$$\lim_{z\to\zeta}(z-\zeta)M_Y(z)=-\frac{p}{(1-p)^2M'(\zeta)}<0,$$
implying that $\zeta$ is a simple pole of $M_Y$.

The fact that $\zeta$ is a positive real pole of $M_Y$ %on the right axis of its region of convergence (i.e., the region of the complex plane where $\mathrm{E}|e^{zY}|<\infty$) 
means that the right tail of the distribution of $Y$ decays to zero exponentially at rate $\zeta$, in the sense that $\ln\Pr(Y>y)\sim-\zeta y$ for large $y$. This is a consequence of a Tauberian theorem proved in Ref.~\cite{Nakagawa2007}. It follows that the right tail of the distribution of $S_T$ exhibits a power law with Pareto exponent $\zeta$: setting $y=\ln s$, we have
$$\lim_{s\to\infty}\frac{\ln \Pr(S_T>s)}{\ln s}=\lim_{y\to\infty}\frac{\ln \Pr(Y>y)}{y}= -\zeta.$$

Eq.~\eqref{eq:BeareToda} appears as Eq.~(10) in Ref.~\cite{ManrubiaZanette1999}. It shows how the Pareto exponent $\zeta$ is uniquely determined by the interaction of the growth rate distribution (through its Laplace transform $M$) and the age distribution (through its parameter $p$). A more general version of Eq.~\eqref{eq:BeareToda} applicable in settings where the growth rates may not be i.i.d.\ but instead satisfy a weaker condition involving Markov modulation has been established in Ref.~\cite{BeareToda-dPL}.

%The connection between Tauberian theorems and generative mechanisms for power laws involving multiplicative growth and exponentially distributed ages was noted in Ref.~

We assumed in the preceding discussion that $T$ has the geometric distribution. This assumption was stronger than necessary; what matters is that the right tail of the distribution of $T$ decays at an exponential rate. In our empirical analysis in Sec.~\ref{empirics} we employ a truncated logistic parametrization of the distribution of $T$, which has an exponentially decaying right tail but in other respects does not resemble the geometric distribution. We will see in Sec.~\ref{empirics6} that Eq.~\eqref{eq:BeareToda} remains valid in this case, with $p$ determined by the rate at which the right tail of the truncated logistic distribution decays exponentially to zero. We may also allow the initial size $S_0$ to be a positive random variable independent of $\{G_t\}_{t=1}^\infty$ and $T$, provided that it satisfies $\mathrm{E}S_0^{\zeta+\epsilon}<\infty$ for some $\epsilon>0$; this can be shown using Breiman's lemma \cite{breiman1965}.

\subsection{\label{theory2}The Susceptible-Infected-Recovered model}

Here we provide a brief discussion of the Susceptible-Infected-Recovered (SIR) model of epidemics \cite{KermackMcKendrick1927} and the extent to which it is consistent with Gibrat's law. In the SIR model, a community (say, a county) consists of a mass of individuals who are either susceptible to an infectious disease (they are neither infected nor have immunity), infected, or immune (possibly because they are vaccinated, infected and recovered, or dead). Individuals meet each other randomly, and conditional on an infected individual meeting a susceptible individual, the disease is transmitted with some probability. Let $\beta>0$ be the rate at which an infected individual meets a person and transmits the disease if susceptible. Let $\gamma>0$ be the rate at which an infected individual recovers or dies. Letting $x,y,z$ be the fractions of susceptible, infected, and recovered individuals in the community (so $x+y+z=1$), the SIR model is described by the system of differential equations
\begin{subequations}\label{eq:xyz}
	\begin{align}
	\dot{x}&=-\beta xy, \label{eq:xyz.x}\\
	\dot{y}&=\beta xy-\gamma y, \label{eq:xyz.y}\\
	\dot{z}&=\gamma y,\label{eq:xyz.z}
	\end{align}
\end{subequations}
where $\dot{x},\dot{y},\dot{z}$ are the rates of change of $x,y,z$.

Although the system of differential equations \eqref{eq:xyz} is nonlinear, it admits an exact analytical solution \cite{HarkoLoboMak2014}. It suffices for our purposes to study the system \eqref{eq:xyz} heuristically at the beginning of the epidemic, where $x\approx 1$ and $y,z$ are small. Setting $x=1$, Eq.~\eqref{eq:xyz.y} becomes $\dot{y}=(\beta-\gamma)y$, and hence $y(t)=y_0e^{(\beta-\gamma)t}$. Integrating Eq.~\eqref{eq:xyz.z}, we obtain
$$z(t)=z_0+\int_0^t\gamma y(s)d s=z_0+y_0\frac{\gamma}{\beta-\gamma}(e^{(\beta-\gamma)t}-1).$$
The cumulative number of cases up to time $t$ is therefore given by
$$c(t)= y(t)+z(t)=z_0+\frac{y_0}{\beta-\gamma}(\beta e^{(\beta-\gamma)t}-\gamma).$$
Assuming that $\beta>\gamma$ (so that there is an epidemic), that time $t$ is neither too large (so that the approximation $x\approx 1$ is valid) nor too small (so that $\gamma\ll \beta e^{(\beta-\gamma)t}$), and that $z_0$ is small relative to $y_0$, it follows that
$$c(t)\approx y_0\frac{\beta}{\beta-\gamma} e^{(\beta-\gamma)t},$$
so cases grow exponentially at rate $g\coloneqq \beta-\gamma>0$. This implies that the growth factor for cases between day $t$ and $t+1$,
\begin{equation}
G_{t+1}=c(t+1)/c(t)\approx e^{g},\label{eq:Gc}
\end{equation}
is independent of the current size $c(t)$. In practice, the transmission rate $\beta$ may fluctuate over time, so it may be plausible to assume that the growth factor $G_{t+1}$ is a random variable independent of the current size $c(t)$, as in Gibrat's law.

The point of the preceding heuristic discussion is that, in the SIR model, the growth of infections may be broadly consistent with Gibrat's law in the early stages of an epidemic. The same cannot be said for the later stages of an epidemic. In the exact analytical solution to the system \eqref{eq:xyz} given in Ref.~\cite{HarkoLoboMak2014}, the growth rate of infections falls as the proportion of the population that is infected or recovered rises, because a smaller proportion of the population remains susceptible to infection. Furthermore, the growth rate of infections may fall as individuals take more precautionary measures such as avoiding crowded spaces. Our empirical findings reported in Sec.~\ref{empirics} are based on data from the onset of COVID-19 in the US in January 2020 until the end of March 2020. At the end of that period the total number of confirmed COVID-19 cases in the US (182,308) was less than 0.06\% of the total population (330 million). We provide evidence in Sec.~\ref{empirics3} that the growth rate of cases remained independent of the number of cases up until the end of March 2020, consistent with Gibrat's law.

\section{\label{empirics}Empirical findings}

\subsection{\label{empirics1}Dataset}

Our dataset consists of the daily numbers of confirmed COVID-19 cases in US counties reported by \emph{The New York Times} \footnote{\url{https://github.com/nytimes/covid-19-data}}, based on reports from state and local health agencies. These numbers are cumulative. We use data from January 21, 2020, when the first case in the US was reported, through to March 31, 2020. There are a total of 3243 counties in the US (including both states and territories). We include the 2121 counties that reported at least one COVID-19 case by March 31 and exclude the remainder. Exceptionally, the dataset combines the five boroughs of New York City (New York, Kings, Queens, Bronx and Richmond counties) into a single unit called New York City.

%Many states and counties in US have introduced mitigation measures such as social distancing and closure of schools, bars, and restaurants starting as early as March 16, 2020. Because there is a time lag between infection and the onset of symptoms (and reporting of cases), we conjecture that it takes about two weeks until these measures show any effect on the virus spread. For this reason, we limit the sample period through the end of March 2020.

\subsection{\label{empirics2}Distribution of COVID-19 cases on March 31}

\begin{figure}
	\includegraphics[width=\linewidth]{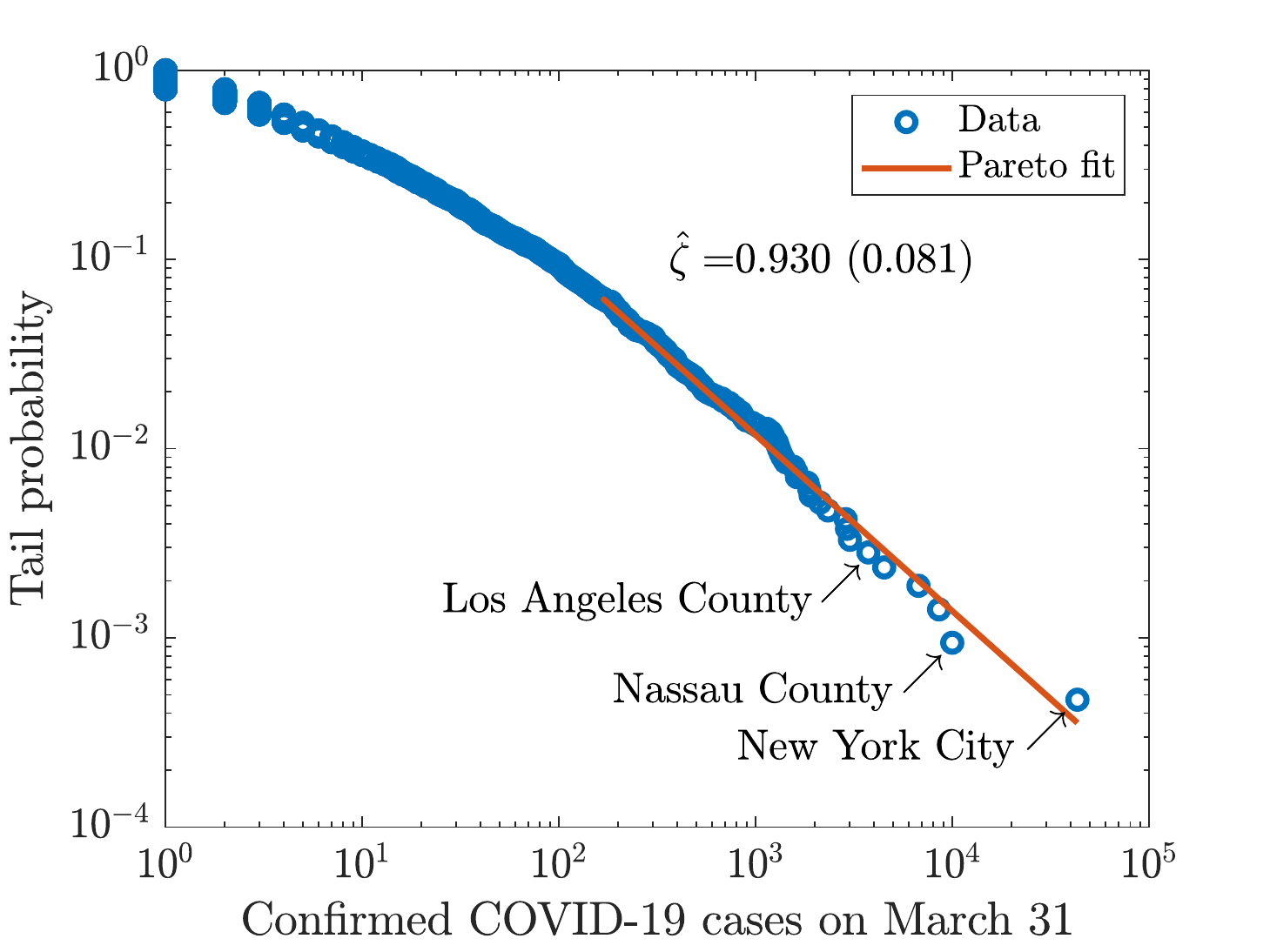}
	\caption{\label{fig:Pareto_Cases}Log-log plot of confirmed COVID-19 cases against tail probabilities across US counties on March 31, 2020. The tail probability of a county is the proportion of all counties matching or exceeding its number of COVID-19 cases. The Pareto fit was obtained by applying the Hill estimator to the top 6.2\% of counties by number of cases. The estimated Pareto exponent is $\hat{\zeta}=0.930$, with a standard error of 0.081.}
\end{figure}

In Fig.~\ref{fig:Pareto_Cases} we plot the number of confirmed COVID-19 cases in each county on March 31 against the corresponding tail probabilities in log-log scale. The tail probability for county $i$ is defined to be the fraction of counties whose number of COVID-19 cases is greater than or equal to that of county $i$. The county with the smallest tail probability, and therefore the highest number of COVID-19 cases, is New York City (though it is in fact an aggregate of five counties, as pointed out in Sec.~\ref{empirics1}). The county with the next highest number of COVID-19 cases is Nassau County, which is located in Long Island and borders Queens County in New York City. The county with the highest population is Los Angeles County, which has the sixth highest number of COVID-19 cases.

Setting aside the two data points for New York City and neighboring Nassau County, the data toward the lower-right of Fig.~\ref{fig:Pareto_Cases} appear to lie roughly on a straight line. This pattern is indicative of a power law in the right tail of the distribution of COVID-19 cases across US counties. To investigate further, we followed the approach recommended in Ref.~\cite{ClausetShaliziNewman2009}. Specifically, we used a version of the Hill estimator \cite{hill1975} to fit a Pareto distribution to the data exceeding a threshold selected using the algorithm described in Ref.~\cite{ClausetYoungGleditsch2007}. Slightly more than 6\% of counties had COVID-19 cases exceeding the selected threshold. The Hill estimate of the Pareto exponent was $\hat{\zeta}=0.930$, with a standard error of 0.081. (Similar results were obtained using a 5\% or 10\% threshold.)
%The p-value for a Kolmogorov-Smirnov test of the adequacy of the Pareto approximation to the data exceeding the selected threshold, computed as described in Ref.~\cite{ClausetYoungGleditsch2007}, was 0.079, indicating statistically acceptable fit. 
The fitted Pareto tail is displayed in Fig.~\ref{fig:Pareto_Cases}, where in log-log scale it appears as a straight line with slope $-\hat{\zeta}$.

The two data points for New York City and Nassau County, which recorded the highest numbers of COVID-19 cases, lie somewhat to the right and to the left of our estimated power law in Fig.~\ref{fig:Pareto_Cases}. In its notes on methodology and definitions, \emph{The New York Times} states that where possible it assigned cases to the county where they were treated, not where they resided \cite{Note1}. This could mean that a significant number of Nassau County residents who were confirmed as having COVID-19 but received treatment in New York City are classified as New York City cases rather than Nassau County cases. If we suppose that one third of Nassau County residents confirmed with COVID-19 were classified as New York City cases, and reassign those cases to Nassau County, then the small circles representing Nassau County and New York City in Fig.~\ref{fig:Pareto_Cases} shift to the right and left respectively, such that both are touching our line of Pareto fit.

\subsection{\label{empirics3}Empirical plausibility of Gibrat's law}

The remainder of our empirical analysis focuses on determining whether the power law with estimated exponent $\hat{\zeta}=0.930$ obtained in Sec.~\ref{empirics2} can be explained by a combination of Gibrat's law and an age distribution with exponential right tail, as described in Sec.~\ref{theory1}. We first assess the empirical plausibility of Gibrat's law as a description of the growth in COVID-19 cases within counties. To this end, for each day $t$ between March 3 and March 30 inclusive, we estimate the cross-sectional regression equation
\begin{equation}
\Delta \ln c_{i,t+1}=\beta_{0t}+\beta_{1t}\ln c_{it}+\beta_{2t}\Delta\ln c_{it}+\beta_{3t}D_{it}+\varepsilon_{it}\label{eq:reg}
\end{equation}
by ordinary least squares, where $c_{it}$ is the number of cases in county $i$ up to day $t$, $\Delta\ln c_{i,t+1}$ is the growth rate in cases in county $i$ between day $t$ and $t+1$, $D_{it}$ is the number of days (inclusive) between day $t$ and the day of the first confirmed case in county $i$, $\varepsilon_{it}$ is the regression residual, and $\beta_{0t},\beta_{1t},\beta_{2t},\beta_{3t}$ are regression coefficients that are potentially time-varying. (Here $\beta_{0t}$ corresponds to the growth rate $g=\beta-\gamma$ in Eq.~\eqref{eq:Gc}.) The estimation of Eq.~\eqref{eq:reg} on day $t$ uses the data for all counties $i$ reporting a positive number of cases ($c_{it}>0$). We estimate Eq.~\eqref{eq:reg} for the 28 days between March 3 and March 30 inclusive because these are the days on which at least 30 counties reported a positive number of cases. The number of counties used in each regression increases from 32 on March 3 to 1940 on March 30.

\begin{figure}
	\subfloat[Intercept ($\beta_{0t}$)\label{subfig-1}]{%
		\includegraphics[width=0.235\textwidth]{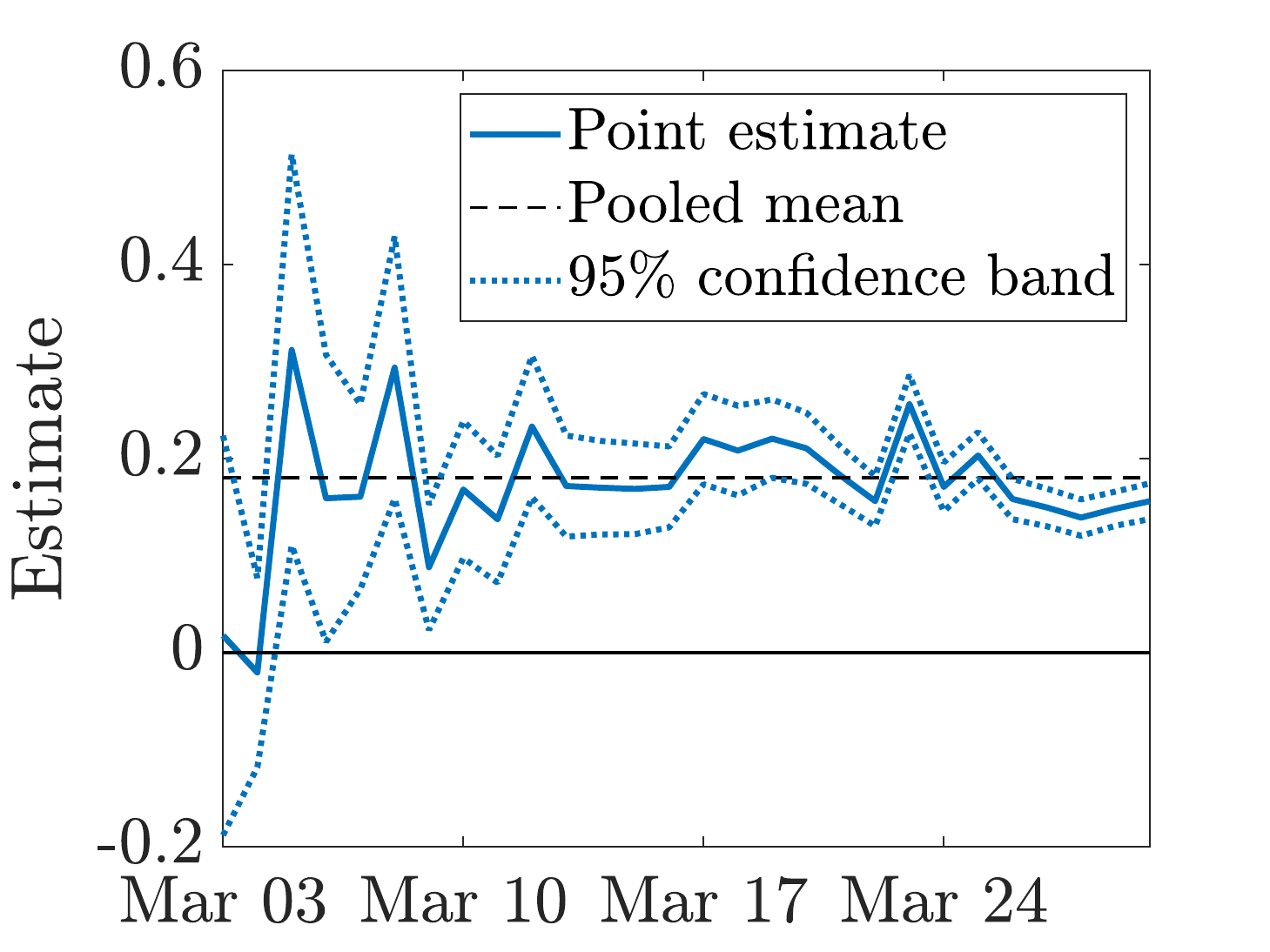}
	}
	\hfill
	\subfloat[Coefficient of log-cases ($\beta_{1t}$)\label{subfig-2}]{%
		\includegraphics[width=0.235\textwidth]{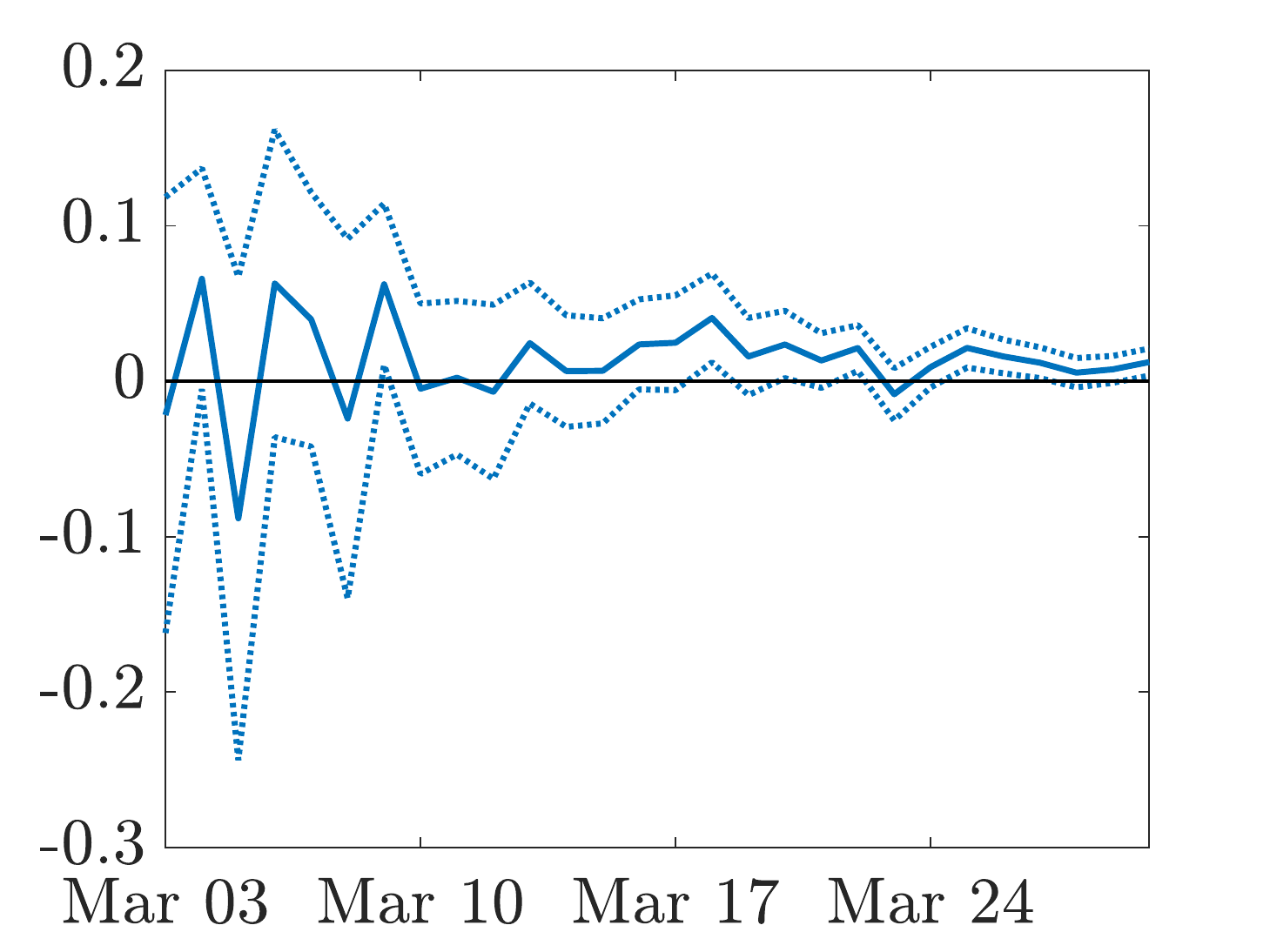}
	}
	\\
	\subfloat[Coefficient of growth rate of cases ($\beta_{2t}$)\label{subfig-3}]{%
		\includegraphics[width=0.235\textwidth]{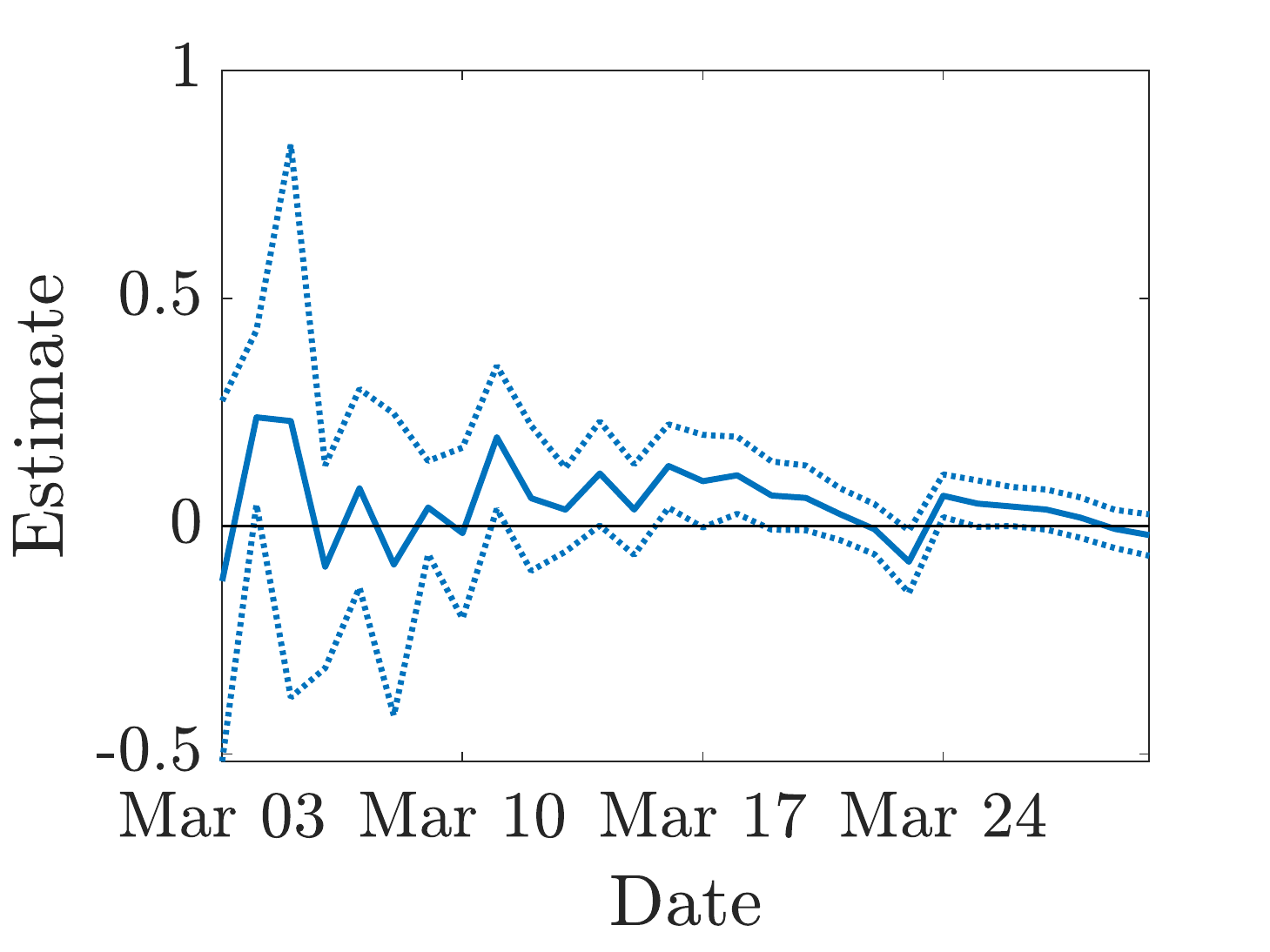}
	}
	\hfill
	\subfloat[Coefficient of days since first confirmed case ($\beta_{3t}$)\label{subfig-4}]{%
		\includegraphics[width=0.235\textwidth]{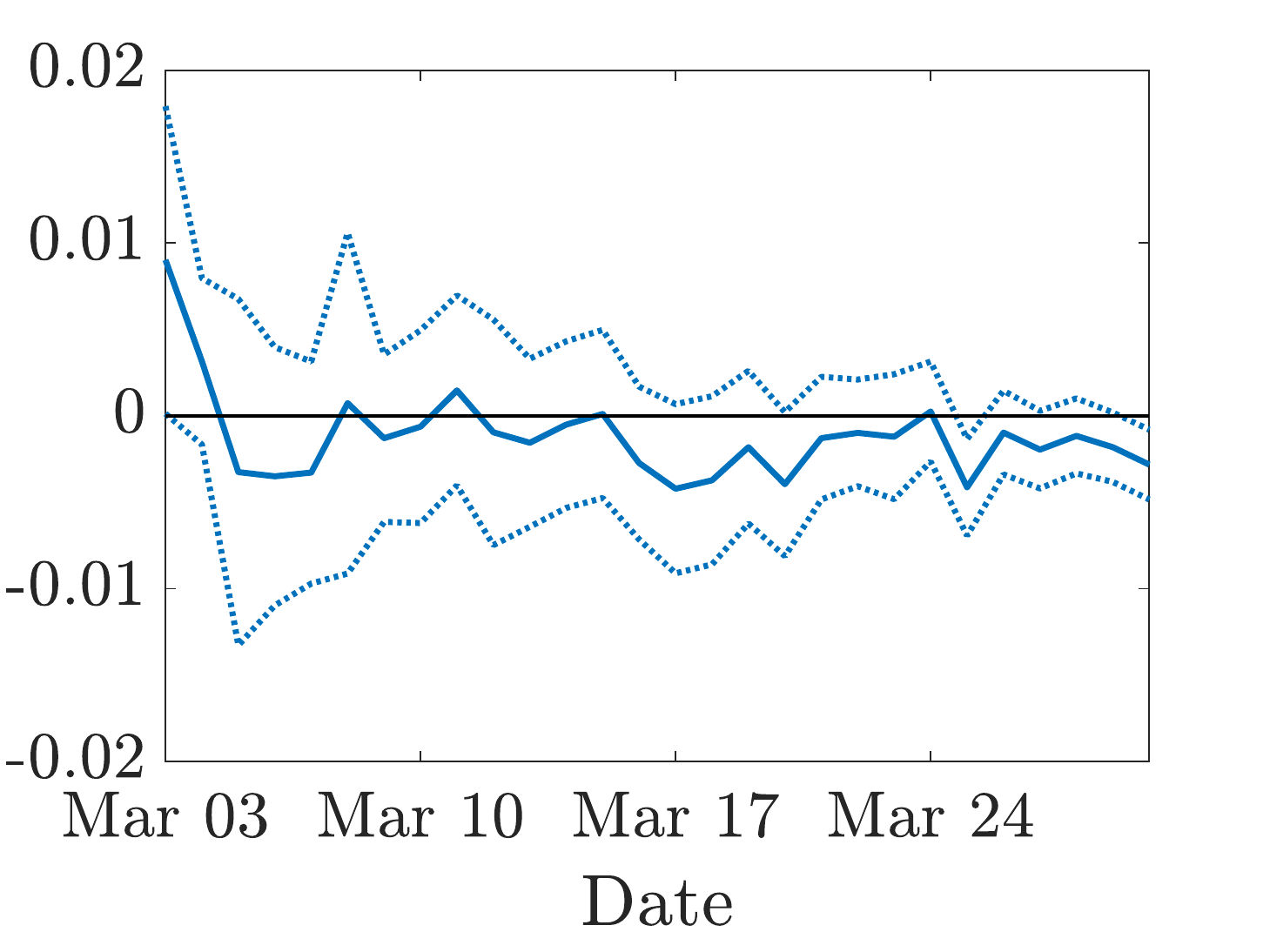}
	}
	\caption{Ordinary least squares estimates of $\beta_{0t},\beta_{1t},\beta_{2t},\beta_{3t}$ in Eq.~\eqref{eq:reg} for the 28 days between March 3 and March 30 inclusive, with 95\% confidence bands. In panel (a) we also display the pooled mean growth rate of 0.180.}
	\label{fig:OLS}
\end{figure}

Fig.~\ref{fig:OLS} displays our estimates of $\beta_{0t},\beta_{1t},\beta_{2t},\beta_{3t}$ in Eq.~\eqref{eq:reg} from March 3 to March 30, with accompanying 95\% confidence bands. In each panel, the confidence bands narrow as we move from left to right, reflecting the fact that the regression sample size increases from 32 to 1940. In panels (b)-(d), we see that the estimates of $\beta_{1t},\beta_{2t},\beta_{3t}$ are close to zero. This is exactly what we would expect under Gibrat's law: the growth rate in cases between days $t$ and $t+1$ ought to be unrelated to the number of cases on day $t$, the growth rate in cases between days $t-1$ and $t$, and the time elapsed since the outbreak.

The estimated parameters are quite stable over time. The estimates of $\beta_{0t}$ displayed in panel (a) indicate that the expected growth rate of cases during March was roughly stable at around 15--20\% per day. The pooled mean growth rate (i.e., the average over all days and counties with at least one reported case) was 18\%, which falls outside the daily 95\% confidence bands on only three days excluding the very end of March, when mitigation efforts may have begun to slow the epidemic. The pooled mean growth rate may be compared to estimates of related parameters obtained in prior research on COVID-19. In the context of the SIR model described in Sec.~\ref{theory2}, the recovery rate $\gamma$ is a biological parameter determined by the virus. In Ref.~\cite{Li_2020} the mean serial interval, which corresponds to $1/\gamma$, is estimated through contact tracing to be 7.5 days. Therefore setting $\gamma=1/7.5=0.133$ and $g=\beta-\gamma=0.180$, our pooled mean growth rate, we estimate the transmission rate $\beta$ to be 0.314. This in turn implies that the reproductive number of COVID-19 (which plays an important role in epidemic dynamics) is $R_0=\beta/\gamma=2.35$, which is close to the estimate of 2.2 reported in Ref.~\cite{Li_2020}. %Although our focus is on the science of power law phenomena, our estimate of the transmission rate may be useful for public health policies since we estimate it from a large dataset.}

%While the fact that $\beta_{1t}=\beta_{2t}=\beta_{3t}=0$ in the regression \eqref{eq:reg} supports Gibrat's law, it only implies that the \emph{conditional mean} of the growth rate of cases is independent of the current size, lagged growth, and days since the outbreak. It is still possible that the conditional growth rate \emph{distribution} depends on these variables, for example through the change in the variance. To address this concern, Figures \ref{fig:pdf_size}--\ref{fig:pdf_days} show the kernel density estimates of the growth rate distribution during the last 14 days in March 2020 conditional on the initial number of cases (size), lagged growth, and days since the outbreak. For each conditioning variable, we pool all growth rates during the last 14 days of March across all counties (with at least 10 cases, to avoid the integer problem) that fall into each quartile sorted by the conditioning variable. While the conditional densities are noisy due to the smaller sample size, they are generally similar, providing further support that the growth rates are independent.

\subsection{\label{empirics4}Distribution of growth rate of confirmed cases}

\begin{figure}[t!]
	\centering
	\includegraphics[width=\linewidth]{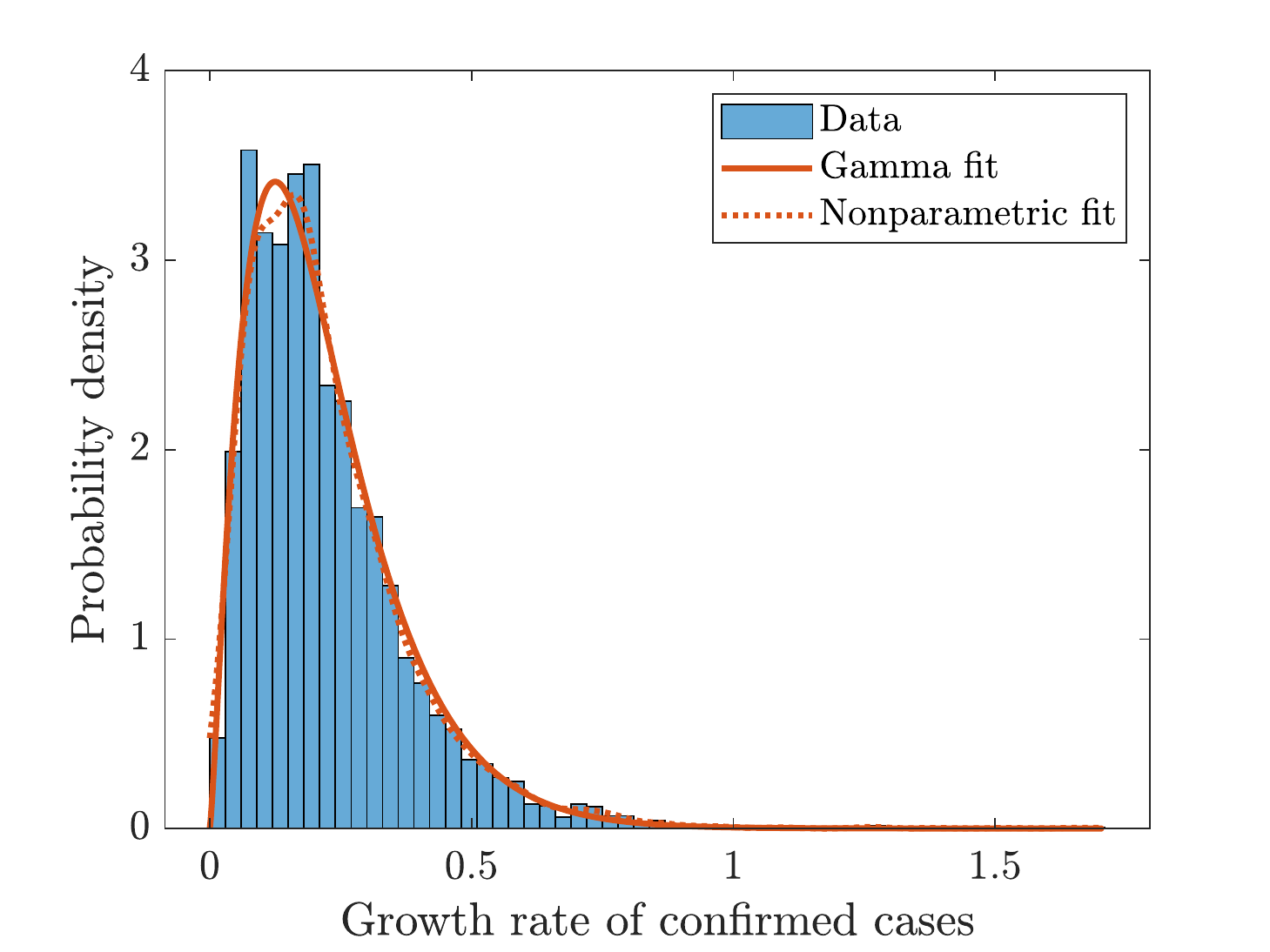}
	\caption{Histogram of growth rates of confirmed COVID-19 cases, using data from all days up to the end of March and all counties with at least 10 confirmed cases and a positive growth rate of cases. The gamma fit was obtained by the method of maximum likelihood. The nonparametric fit was obtained by Gaussian kernel smoothing.}\label{fig:hist_Growth}
\end{figure}

In Fig.~\ref{fig:hist_Growth} we display a histogram of the growth rates of confirmed COVID-19 cases, obtained by pooling our data across all days $t$ up to March 30 and all counties $i$ with at least 10 confirmed cases ($c_{it}\geq10$) and a positive growth rate ($\Delta\ln c_{i,t+1}>0$). Overlaying the histogram we plot a gamma distribution fit to our data by the method of maximum likelihood. The gamma distribution has density
\begin{equation}
f(x)=\frac{\beta^\alpha}{\Gamma(\alpha)}x^{\alpha-1}e^{-\beta x},\quad x>0,\label{eq:gamma}
\end{equation}
where $\alpha,\beta>0$ are called the shape and rate parameters. The maximum likelihood parameter estimates are  $\hat{\alpha}=2.30$ and  $\hat{\beta}=10.4$. The fit of the gamma distribution appears to be excellent, particularly in the right tail of the data. It closely matches a nonparametric estimate obtained using Gaussian kernel smoothing, which we plot alongside it in Fig.~\ref{fig:hist_Growth}.

Our data up to March 30 include a total of 5687 day-county pairs with at least 10 confirmed cases ($c_{it}\geq10$). Of those, 725 day-county pairs had a zero growth rate of cases ($\Delta\ln c_{i,t+1}=0$), and were therefore excluded from the computation of the histogram in Fig.~\ref{fig:hist_Growth} and corresponding gamma fit. An observed growth rate could be zero because there were indeed no new cases, or because the data were not updated on a particular day. The proportion of growth rates observed to be zero is $\pi=725/5687=0.128$, which is substantial. For this reason, in the analysis in Sec.~\ref{empirics6}, we model the distribution of the growth rate of confirmed cases as a mixture of our maximum likelihood estimate of the gamma distribution plotted in Fig.~\ref{fig:hist_Growth} and a point mass at zero, with proportions $1-\pi$ and $\pi$ respectively.

\subsection{\label{empirics5}Distribution of days since first confirmed case}

\begin{figure}
	\centering
	\includegraphics[width=\linewidth]{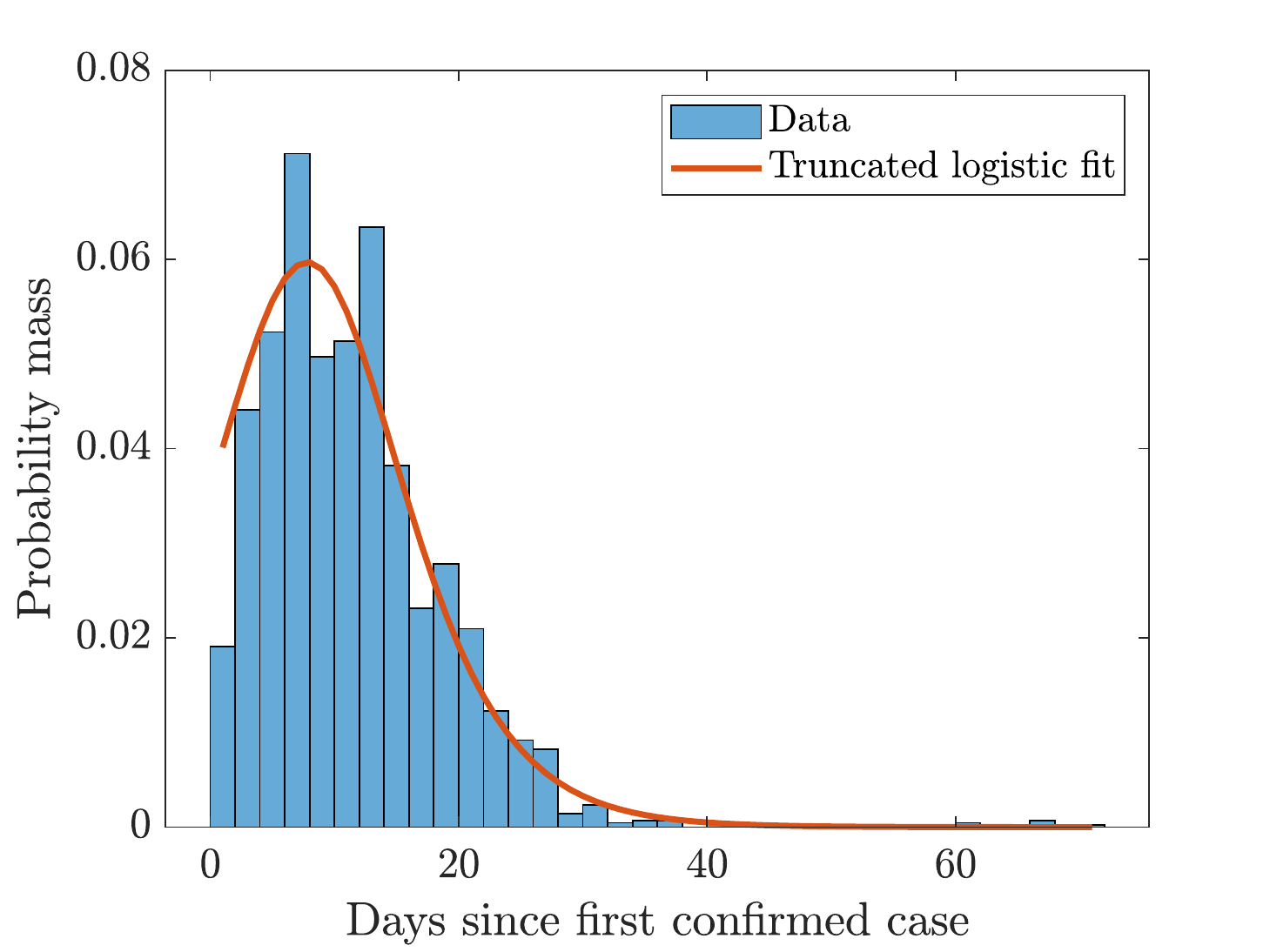}
	\caption{Histogram of days-since-outbreak on March 31, using data from all counties reporting at least one confirmed COVID-19 case by March 31. The truncated logistic fit was obtained by the method of maximum likelihood.}\label{fig:hist_Days}
\end{figure}

In Fig.~\ref{fig:hist_Days} we display a histogram of the number of days (inclusive) between the day of the first confirmed case of COVID-19 in a county, and March 31. The histogram is computed from the 2121 counties in our dataset that reported at least one confirmed case by March 31.

As discussed earlier, the combination of Gibrat's law with an exponential age distribution has been widely used as a generative mechanism for power laws. It is apparent from the histogram in Fig.~\ref{fig:hist_Days}, however, that the exponential distribution (or its discrete counterpart, the geometric distribution) cannot provide an acceptable approximation to the age distribution we observe in our data. The problem is that the density of the exponential distribution is monotonically decreasing, whereas the histogram in Fig.~\ref{fig:hist_Days} is roughly hump shaped.

In nature, an exponential distribution of ages arises when a population grows exponentially over time, as in the Yule model of speciation discussed in Refs.~\cite{reed2001,reed2002}. By analogy, we may expect to see an exponential distribution of days-since-outbreak in our data if the number of counties that have reported at least one confirmed COVID-19 case is growing exponentially over time. The analogy fails because there are only 3243 counties in the US, so that exponential growth cannot be maintained. Once COVID-19 has spread to a substantial proportion of counties, the rate of growth in the number of counties reporting at least one case ought to fall, eventually vanishing as saturation is approached. Given that nearly two thirds of US counties had reported at least one confirmed case by March 31, we would expect the number of newly infected counties to be declining by this time. This explains the hump shape in the histogram in Fig.~\ref{fig:hist_Days}.

The logistic function was introduced in the 19th century as a model of population growth that commences at an exponential rate but tapers off as a saturation point is approached due to competition for resources \cite{cramer2003}. In the SIR model discussed in Sec.~\ref{theory2}, in the absence of recovery (so that $\gamma=0$ in Eq.~\eqref{eq:xyz.y}), a logistic function describes the growth of the infected population over time, and the distribution of time-since-infection for the infected population at any given time is the truncated logistic distribution. (Truncation is always necessary because saturation is not reached in finite time.) By analogy, when considering the spread of an infection across a population of counties, we might expect the truncated logistic distribution to well-approximate the distribution of days-since-outbreak across counties at any given time. We therefore truncate a discrete version of the logistic distribution introduced in Ref.~\cite{ChakrabortyChakravarty2016}. 
%Based on the discussion in the previous paragraph, it seems reasonable to use a truncated logistic distribution to model the distribution of days-since-outbreak on March 31. In fact, if we assume COVID-19 is randomly transmitted from counties with reported cases to those without yet, the age distribution becomes truncated logistic. Truncation is necessary because saturation of counties has not been reached by March 31. We truncate a discrete version of the logistic distribution introduced in Ref.~\cite{ChakrabortyChakravarty2016}. 
Without truncation, for any integer $n$, the discrete logistic distribution has probability mass
\begin{equation}
\Pr(T=n)=\frac{(1-q)q^{n-\mu}}{(1+q^{n-\mu})(1+q^{n-\mu+1})},\label{eq:logistic}
\end{equation}
where $q\in(0,1)$ is a parameter determining the rate of exponential decay in the tails, and $\mu$ is a location parameter. After truncating all mass on nonpositive integers and rescaling so that the remaining mass sums to one, the probability mass assigned to each integer $n\geq1$ is
\begin{equation}
\Pr(T=n)=\frac{(1+\lambda)(1-q)q^{n-1}}{(1+\lambda q^{n-1})(1+\lambda q^n)},\label{eq:trunclogistic}
\end{equation}
where we have reparametrized the distribution in terms of $q$ and $\lambda=q^{1-\mu}$, the latter equal to the ratio of probability masses included and excluded by truncation.

Overlaying the histogram in Fig.~\ref{fig:hist_Days} we plot a truncated logistic distribution fit to our data by the method of maximum likelihood. The maximum likelihood parameter estimates are $\hat{q}=0.825$ and $\hat{\lambda}=4.06$. The fit captures the general shape of the histogram reasonably well, particularly toward the right tail, which is the more important region for our purposes.

\subsection{\label{empirics6}Implied Pareto exponent}

\begin{figure}[t!]
	\centering
	\begin{tikzpicture}[scale=1.8]
	\draw[black, thick, ->] (0,0) -- (0,2.75);
	\draw[black, thick, ->] (0,0) -- (3,0);
	\draw[black, dotted] (0,1.2115) -- (3,1.2115);
	\draw[black, dotted] (0.9361,0) -- (0.9361,1.2115);
	\draw[black,domain=0:3, ->]  plot(\x, {725/5687+(1-725/5687)*(1-\x/10.3792)^(-2.2963)});
	\node[below] at (0.9361,0) {\footnotesize \(\zeta\)};
	\node[below] at (3,0) {\footnotesize \(z\)};
	\node[above] at (3,{725/5687+(1-725/5687)*(1-3/10.3792)^(-2.2963)}) {\footnotesize \(M(z)\)};
	\node[below left] at (0,0) {\footnotesize \(0\)};
	\node[left] at (0,1) {\footnotesize \(1\)};
	\node[left] at (0,1.2115) {\footnotesize \(1/q\)};
	\end{tikzpicture}
	\caption{The Pareto exponent $\zeta$ is the unique positive real $z$ at which the Laplace transform $M(z)$ is equal to $1/q$.}\label{fig:BeareToda}
\end{figure}
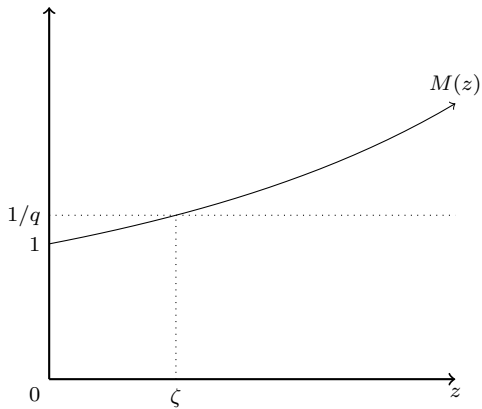

In Sec.~\ref{theory1} we discussed how the combination of Gibrat's law and an exponential age distribution can generate a power law, with Pareto exponent $\zeta$ solving Eq.~\eqref{eq:BeareToda}. It remains for us to determine the Pareto exponent thus obtained when the distributions of growth rates and ages are as estimated in Secs.\ \ref{empirics4} and \ref{empirics5}. A complicating factor is that our age distribution is not exactly exponential, but instead belongs to the family of truncated logistic distributions defined by Eq.~\eqref{eq:trunclogistic}. This leads us to replace Eq.~\eqref{eq:MY} with
\begin{equation}\label{eq:MYlog}
M_Y(z)=\sum_{n=1}^\infty \frac{(1+\lambda)(1-q)q^{n-1}}{(1+\lambda q^{n-1})(1+\lambda q^n)}M(z)^n,
\end{equation}
valid for all positive real $z$ such that $qM(z)<1$. Define
$$r_n=[(1+\lambda q^{n-1})(1+\lambda q^n)]^{-1}-1,$$
and let $p=1-q$. It is straightforward to show that $|r_n|\leq(1+\lambda)^2q^{n-1}$, so we may rewrite Eq.~\eqref{eq:MYlog} as
$$M_Y(z)=(1+\lambda)p\left[\frac{M(z)}{1-qM(z)}+\sum_{n=1}^\infty q^{n-1}r_nM(z)^n\right],$$
where the first term in square brackets has a pole at the unique $\zeta\in(0,\eta)$ solving Eq.~\eqref{eq:BeareToda} and the second term in square brackets is analytic in a neighborhood of $\zeta$. This shows that $\zeta$ is a positive real pole of $M_Y$ and so, as in Sec.~\ref{theory1}, we deduce from the Tauberian theorem proved in Ref.~\cite{Nakagawa2007} that the right tail of the distribution of $S_T$ exhibits a power law with Pareto exponent $\zeta$. Fig.~\ref{fig:BeareToda} displays visually how $\zeta$ is determined by the parameter $q$ and Laplace transform $M(z)$, with our empirical estimates for $q$ and $M(z)$.

%\begin{comment}
%\end{comment}

Our estimate of the distribution of the growth rate of confirmed COVID-19 cases (i.e., the distribution of $X_t$) reported in Sec.~\ref{empirics4} was a mixture of a point mass at zero and a gamma distribution with weights $\pi=0.128$ and $1-\pi$ respectively. The particular form of this distribution allows us to obtain the solution $\zeta$ to Eq.~\eqref{eq:BeareToda} in analytic form. Specifically, the Laplace transform of the distribution of $X_t$ is given for real $z<\beta$ by
\begin{align}
M(z)&=\pi+(1-\pi)\int_0^\infty e^{zx}\frac{\beta^\alpha}{\Gamma(\alpha)}x^{\alpha-1}e^{-\beta x}d x\notag \\
&=\pi+(1-\pi)\left(1-\frac{z}{\beta}\right)^{-\alpha},\label{eq:MX}
\end{align}
and so the unique solution to Eq.~\eqref{eq:BeareToda} is
\begin{equation}
\zeta=\beta\left[1-\left(\frac{1-\pi}{q^{-1}-\pi}\right)^{1/\alpha}\right].\label{eq:impliedzeta}
\end{equation}
Substituting the empirical estimates $\pi=0.128$, $\hat{\alpha}=2.30$, $\hat{\beta}=10.4$ and $\hat{q}=0.825$ into Eq.~\eqref{eq:impliedzeta}, we obtain the implied Pareto exponent $\zeta=0.936$, which is nearly exactly equal to (and well within the 95\% confidence interval of) the estimate $\hat{\zeta}=0.930$ reported in Sec.~\ref{empirics2}. (If we use the nonparametric distribution in Fig.~\ref{fig:hist_Growth} in place of the gamma distribution then the implied Pareto exponent is $\zeta=0.928$.) Thus our simple model involving Gibrat's law generates precisely the power law we observe in the distribution of COVID-19 cases across US counties at the end of March.

\section{\label{conclusion}Final remarks}

We conclude with some brief remarks in nontechnical language to summarize the primary import of our results to policy-makers dealing with the COVID-19 epidemic, or to historians seeking to understand the early weeks of the COVID-19 epidemic in the US. An empirical feature of the distribution of COVID-19 cases across US counties at the end of March 2020 is that case loads are dramatically higher in some counties than in others. That is, the distribution of COVID-19 cases across US counties at the end of March has a heavy right tail. While it may be natural to look for county-specific characteristics to explain why this is the case, our results indicate that this is not necessary. The very high case loads observed in some counties are accurately predicted by a simple empirically calibrated model combining (i) random multiplicative growth within each county, and (ii) variation across counties in the duration of the spread of COVID-19. There is no need to attribute the highest case loads to other idiosyncratic factors. New York City, where the number of confirmed cases at the end of March substantially exceeds our model's prediction, may be an exception.

\providecommand{\noopsort}[1]{}\providecommand{\singleletter}[1]{#1}%

\end{document}